# Magnetization distribution and domain wall dynamics in nanotube with surface anisotropy

**N A Usov[1,2] and O.N. Serebryakova[1,2]**

[1]National University of Science and Technology «MISIS», 119049, Moscow, Russia
[2]Pushkov Institute of Terrestrial Magnetism, Ionosphere and Radio Wave Propagation, Russian Academy of Sciences, (IZMIRAN) 142190, Troitsk, Moscow, Russia

The period of magnetization oscillations that occur near the surface of a nanotube or nanowire under the influence of surface magnetic anisotropy is determined by means of numerical simulation as a function of nanowire geometry and material parameters. The hopping mode is observed for stationary movement of a head-to-head domain wall along nanowire axis in applied magnetic field. The average speed of the domain wall in the hopping mode is found to be several times less than the stationary velocity of the wall in the absence of surface anisotropy.

Surface magnetic anisotropy[1], which can exist at a boundary of a ferromagnet with vacuum or a non-magnetic material may have a significant influence on magnetic properties of thin magnetic films[2-5] and nanoparticles[6,7]. In this paper we study the effect of surface anisotropy on the properties of magnetic nanowires and nanotubes[8-10], which are promising for various applications, in particular for the development of the racetrack memory[11]. It is shown by means of numerical simulation that periodic perturbation of the magnetization occurs near the surface of a nanotube or nanowire when the surface anisotropy constant is negative and sufficiently high in absolute value. The surface magnetization distributions obtained resemble a domain structure. However, the usual domain structure penetrates deep inside the ferromagnetic sample, while the magnetization perturbations caused by the presence of surface anisotropy decay rapidly away from the sample surface. Nevertheless, it significantly affects the dynamics of a head-to-head domain wall propagating along the nanotube or nanowire axis.

The energy density of the surface magnetic anisotropy is given by[4] $w = K_s(\boldsymbol{\alpha n})^2$, where $K_s$ is the surface anisotropy constant, $\boldsymbol{\alpha}$ is the unit magnetization vector, $\boldsymbol{n}$ being the unit vector normal to the sample surface. The boundary conditions for the vector $\boldsymbol{\alpha} = (\alpha_\rho, \alpha_\varphi, \alpha_z)$ at the sample surface, $\rho = R$, in cylindrical coordinates ($\rho$, $\varphi$, $z$) are as follows

$$C\frac{\partial\alpha_\rho}{\partial\rho} = 2K_s\alpha_\rho\left(\alpha_\rho^2 - 1\right); \quad C\frac{\partial\alpha_\varphi}{\partial\rho} = 2K_s\alpha_\rho^2\alpha_\varphi;$$

$$C\frac{\partial\alpha_z}{\partial\rho} = 2K_s\alpha_\rho^2\alpha_z, \qquad (1)$$

where $C$ is the exchange constant. The magnetization distribution in the sample volume is determined by the stationary Landau-Lifshitz equation $\left[\vec{\alpha}, \vec{H}_{ef}\right] = 0$. For a cylindrical specimen having axial symmetry the components of the total effective magnetic field are

$$H_{ef,\rho} = \frac{C}{M_s}\left\{\frac{1}{\rho}\frac{\partial}{\partial\rho}\left(\rho\frac{\partial\alpha_\rho}{\partial\rho}\right) - \frac{\alpha_\rho}{\rho^2} + \frac{\partial^2\alpha_\rho}{\partial z^2}\right\} + H'_\rho$$

$$H_{ef,\varphi} = \frac{C}{M_s}\left\{\frac{1}{\rho}\frac{\partial}{\partial\rho}\left(\rho\frac{\partial\alpha_\varphi}{\partial\rho}\right) - \frac{\alpha_\varphi}{\rho^2} + \frac{\partial^2\alpha_\varphi}{\partial z^2}\right\}$$

$$H_{ef,z} = \frac{C}{M_s}\left\{\frac{1}{\rho}\frac{\partial}{\partial\rho}\left(\rho\frac{\partial\alpha_z}{\partial\rho}\right) + \frac{\partial^2\alpha_z}{\partial z^2}\right\} - H_a\alpha_z + H'_z \qquad (2)$$

Here $M_s$ is the saturation magnetization, $H_a = 2K/M_s$ is the anisotropy field in the sample volume, $K$ being the magnetic anisotropy constant in the bulk, $H'_\rho$ and $H'_z$ are the components of the demagnetizing field. The easy anisotropy axis in the bulk is parallel or perpendicular to the nanowire axis for $K > 0$ and $K < 0$, respectively.

The numerical solution of Eqs (1) - (2) is obtained for magnetic nanowires and nanotubes of submicron radius, $R \le 0.5$ μm, with saturation magnetization $M_s$ = 350 – 600 emu/cm$^3$, bulk anisotropy constant $|K| = 10^3 - 10^5$ erg/cm$^3$, and exchange constant $C = 2\times10^{-6}$ erg/cm. Due to axial symmetry of the samples, the calculations are carried out using two-dimensional numerical scheme[12]. The size of the toroidal numerical cell, $b_z = b_r$ = 2 - 4 nm, is sufficiently small with respect to the exchange length, $\sqrt{C}/M_s$. The periodic magnetization perturbation at the sample surface occurs suddenly with a finite amplitude when surface anisotropy constant $K_s$ is negative, and its absolute value exceeds certain critical value $K_s^*$. For magnetically soft samples, $K < M_s^2$, the $K_s^*$ value is practically independent of the bulk anisotropy constant $K$, but essentially depends on the saturation magnetization $M_s$.

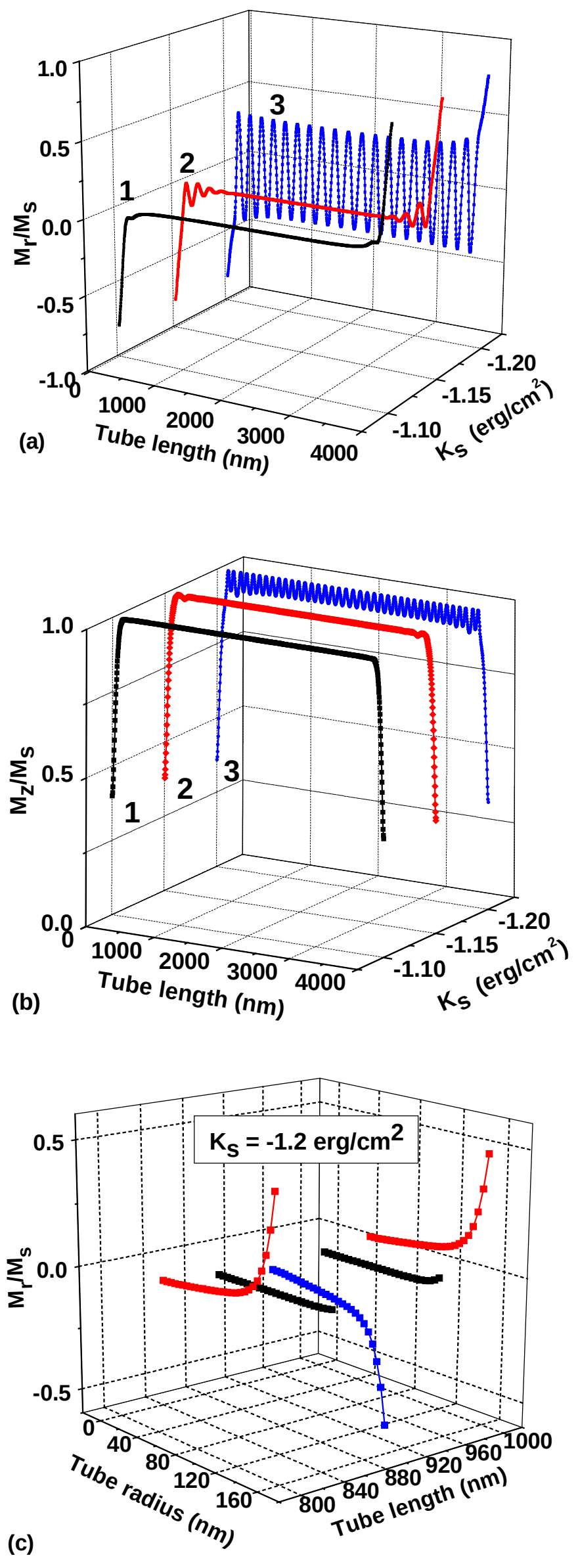

Fig. 1. Radial $\alpha_\rho(R,z)$ (a) and longitudinal $\alpha_z(R,z)$ (b) components of the unit magnetization vector at the surface of a nanotube with $K > 0$ at various $K_s$ values; (c) the behavior of the radial component $\alpha_\rho(\rho,z)$ at $K_s = K_s^*$ within one oscillation period, $\Delta z = l_z$.

Figs. 1a and 1b show the unit magnetization vector components, $\alpha_\rho(R,z)$ and $\alpha_z(R,z)$, at the nanotube surface, $\rho = R$, depending on the value of the surface anisotropy constant. The nanotube length is $L_z = 4000$ nm, the inner and outer radii are $R_1 = 48$ and $R = 160$ nm, respectively, the magnetic parameters equal $K = 10^5$ erg/cm$^3$ and $M_s = 400$ emu/cm$^3$. For values of $|K_s| \le 1.15$ erg/cm$^2$ (see curves 1 and 2 in Figs. 1a, 1b) a sufficiently long nanotube with $K > 0$ is uniformly magnetized along its axis, $\alpha_z = 1$, with the exception of small areas $\Delta z \sim R << L_z$ near the tube ends, where a strong demagnetizing field exists. In these areas the magnetization is twisted along azimuthal and radial directions, while in the central part of the nanotube the components $\alpha_\rho$ and $\alpha_\varphi$ are negligibly small. However, if the surface anisotropy constant reaches a critical value, $K^*_s = -1.2$ erg/cm$^2$, a periodic perturbation of the radial magnetization component $\alpha_\rho$ near the tube surface develops. It has a period $l_z = 204$ nm along the tube axis (see curve 3 in Fig, 1a). The corresponding oscillations of the longitudinal magnetization component, $\alpha_z = \sqrt{1-\alpha_\rho^2}$, have nearly double period (curve 3 in Fig. 1b). On the other hand, in the central part of the nanotube the azimuthal component is still close to zero, $\alpha_\varphi \approx 0$. The reason for the magnetization oscillations is the same as that for the domain structure formation. The deviation of the unit magnetization vector in the radial direction under the influence of the surface anisotropy leads to appearance of the surface magnetic charge, which significantly increases the magneto static energy of the sample. Periodic alternating oscillation of the radial magnetization component turns average surface magnetic charge to zero. As a result, the magneto static energy of the sample reduces substantially. While the exchange energy of the nanotube increases, the total energy decreases. However, unlike the usual domain structure, the magnetization perturbation exists only in the surface region of the nanotube. As Fig. 1c shows, the amplitude of the $\alpha_\rho$ component rapidly decreases to zero with decreasing of the radial coordinate deep into the nanotube volume.

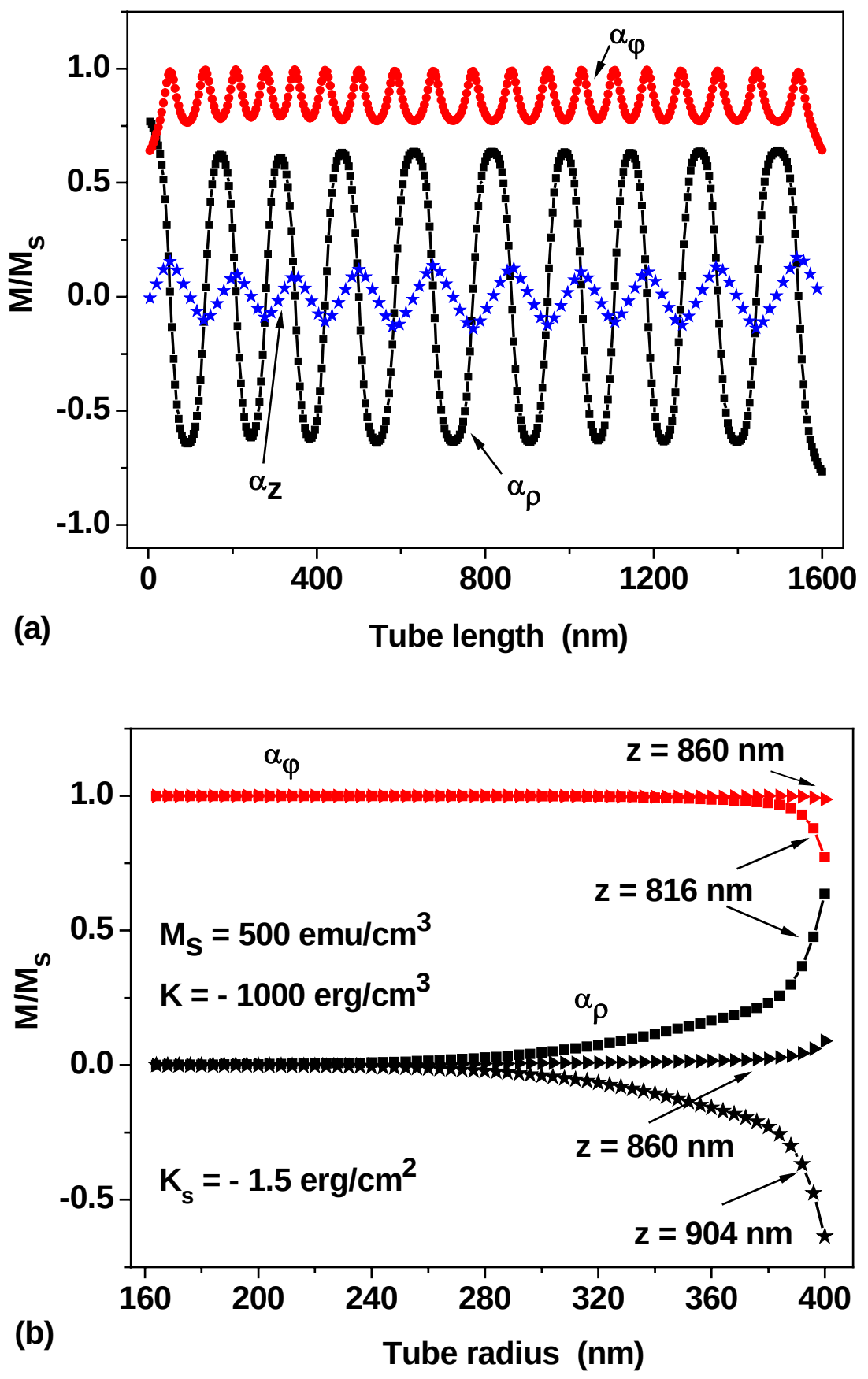

Fig. 2. a) The unit magnetization vector components at the surface of the magnetic nanotube with $K < 0$ for $K_s = -1.5$ erg/cm$^2$; b) radial dependence of $\alpha_\rho$ and $\alpha_\varphi$ components in various cross- sections of the nanotube, at $z = 816$, 860 and 904 nm, respectively.

Similar results were obtained for the nanotubes with $K < 0$. At $|K_s| < |K_s^*|$ the unit magnetization vector of the nanotube has azimuthal direction, $\alpha_\varphi = \pm 1$. Fig. 2a shows a periodic perturbation of the magnetization that arise at the surface of the nanotube with $K = -10^3$ erg/cm$^3$ and $M_s = 500$ emu/cm$^3$ at $K_s = -1.5$ erg/cm$^2$, that is greater in absolute value than the critical value $K_s^* = -1.35$ erg/cm$^2$. The tube has outer and inner radii $R_1 = 160$ and $R = 400$ nm, respectively. The length of the nanotube, $L_z = 1600$ nm, is chosen shorter than in Fig. 1 as for a nanotube with $K < 0$ the magnetization deviations near the tube ends have a relatively small size. Interestingly, in contrast to nanotubes with $K > 0$, where the component $\alpha_\varphi \approx 0$ far from the tube ends, for the present case all magnetization components experience oscillations, the oscillation period of the $\alpha_\rho$ component being $l_z = 176$ nm. In Fig. 2b the radial dependence of the $\alpha_\rho$ and $\alpha_\varphi$ components are shown in different cross-sections of the nanotube along its length within a half-period of the oscillations. As Fig. 2b shows, when $K < 0$ the surface magnetization deviation also decreases rapidly to zero away from the nanotube surface. Therefore, the results for the nanotubes shown in Figs. 1 and 2 are also valid for nanowires with similar dimensions and material parameters.

Fig. 3 shows the dependence of the oscillation period $l_z$ as a function of the saturation magnetization at a corresponding critical value $K_s^*$ for both types of nanotubes, $K > 0$ and $K < 0$. As inset in Fig. 3 shows, the absolute value $|K_s^*|$ increases with increasing of the tube saturation magnetization, whereas the oscillation period $l_z$ decreases. For samples of soft magnetic type, $K < M_s^2$, no noticeable dependence of the $l_z$ and $K_s^*$ quantities on the bulk anisotropy constant $K$ is obtained. The critical values of the surface anisotropy constant obtained numerically, $K^*_s \approx -1$ erg/cm$^2$, are close to that ones observed experimentally in thin magnetic films[2,4]. The calculations presented in Fig. 3 are carried out for the nanotubes with inner and outer radii $R_1 = 128$ and $R = 160$ nm, respectively. The length of the nanotube with $K > 0$ was $L_z = 4000$ nm, while for the nanotube with $K < 0$ it equals $L_z = 1600$ nm. Similar calculations for nanotubes of other sizes reveal only weak dependence of $l_z$ and $K_s^*$ values on the outer radius $R$ of the nanotube, while the outer radius of nanotubes with $K > 0$ has been changed by more than two times, and for nanotubes with $K > 0$ by more than 8 times, respectively.

Thus, the period of the surface magnetization oscillations occurring in the nanotubes of submicron diameter is determined mainly by $K_s$ and $M_s$ values, being virtually independent of the outer sample radius. This is because the amplitude of the surface perturbations decreases rapidly into the sample, as Figs. 1c and 2b show. This means that the surface of the nanotube can be considered locally flat when the outer tube radius $R > 150$ - $200$ nm.

If this conclusion is correct, the period of the magnetization perturbations at the surface of amorphous ferromagnetic microwires with the radius $R = 5$ - $10$ μm caused by the presence of surface magnetic anisotropy must be small compared to their radius.

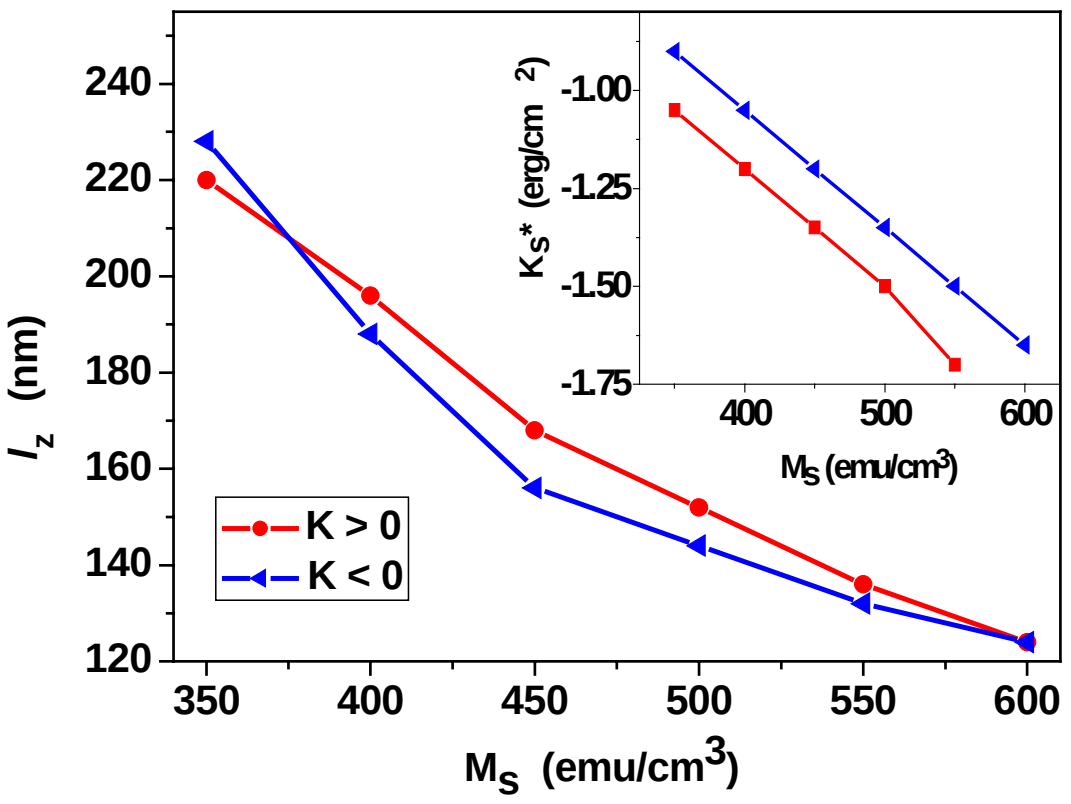

Fig. 3. The period $l_z(M_s)$ of magnetization oscillation along the nanotube axis at the critical value $K_s = K_s^*$ for samples with $K > 0$ and $K < 0$, respectively. Inset shows the $K_s^*(M_s)$ dependence.

Note that such a small-scale magnetization ripples have been observed using magnetic force microscope at the surface of amorphous ferromagnetic microwires in several studies[13,14]. The appearance of the surface magnetic structures with a period small compared with the microwire radius may indicate the presence of the surface magnetic anisotropy.

Since magnetization perturbation exists only in the surface region of the nanotube, it only slightly reduces the total magnetic moment of the tube with $K > 0$. However, the structure of a head-to-head domain wall in magnetic nanotube changes substantially in the presence of surface magnetic anisotropy. Fig. 4a shows the magnetization distribution at the surface of a nanotube with $K > 0$ in the case when there is a head-to-head domain wall within the tube with magnetic parameters $K = 10^5$ erg/cm$^3$, $M_s = 450$ emu/cm$^3$, and $K_s = -1.4$ erg/cm$^2$. The tube has inner and outer radii $R_1 = 96$ and $R = 128$ nm, respectively, and the length $L_z = 4800$ nm. It was shown[12] that in a nanotube with a longitudinal bulk anisotropy in the absence of surface anisotropy the radial magnetization component for a head-to-head domain wall is nonzero only near the center of the domain wall. As Fig. 4a shows, due to the presence of the surface anisotropy there are magnetization oscillations of the $\alpha_\rho$ component near the tube surface with a period $l_z \approx 270$ nm. One can see in Fig. 4a that in the absence of an external magnetic field, the center of the head-to-head domain wall, which corresponds to the maximum of the $\alpha_\varphi$ component, coincides with one of the maxima of the radial magnetization component $\alpha_\rho$. The calculations show that the maxima of the $\alpha_\rho$ component are the pinning centers for the head-to-head domain wall. When a high enough external magnetic field, $H_z = 50$ Oe, is applied along the nanotube axis, after a short transient period of time a stationary hopping motion of the head-to-head domain wall develops, as shown in Fig. 4b. In the hopping mode, the domain wall jumps from one pinning center to the next one. The domain wall dynamics is calculated by means of the Landau-Lifshitz-Gilbert equation[12], the dimensionless magnetic damping constant is assumed to be $\kappa = 0.3$. The speed of the domain wall is calculated as the time derivative of the average magnetic moment of the nanotube along its axis. The maxima of the wall velocity in Fig. 4b correspond to the times when there is a jump of the domain wall center from one pinning center to another. In the absence of surface anisotropy the stationary velocity of the head-to-head domain wall in the same nanotube at given conditions is calculated to be $V = 196$ m/s. However, as Fig. 4b demonstrates, the average velocity of the head-to-head domain wall reduces up to $V \approx 50$ m/s under the influence of the surface anisotropy of appreciable value. Therefore, the surface magnetic anisotropy may significantly decrease the average velocity of a head-to-head domain wall in nanowires and nanotubes with the longitudinal magnetic anisotropy. This fact is important for various applications where fast propagation of the domain wall along the sample is necessary.

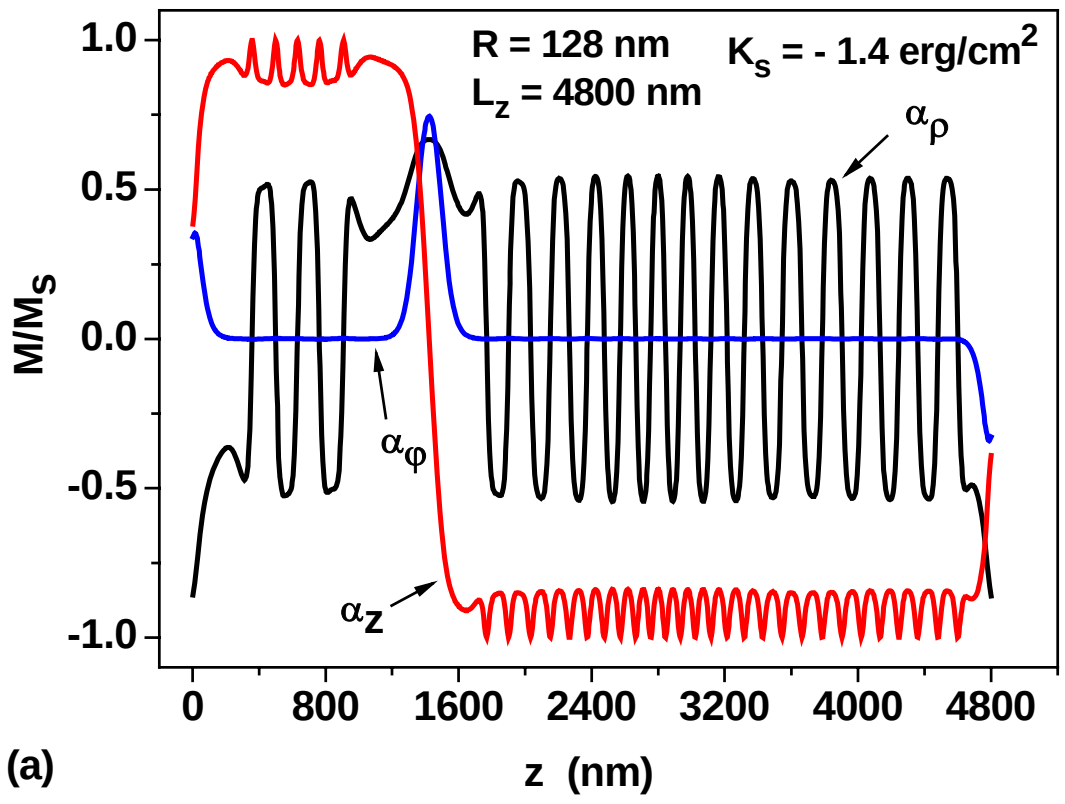

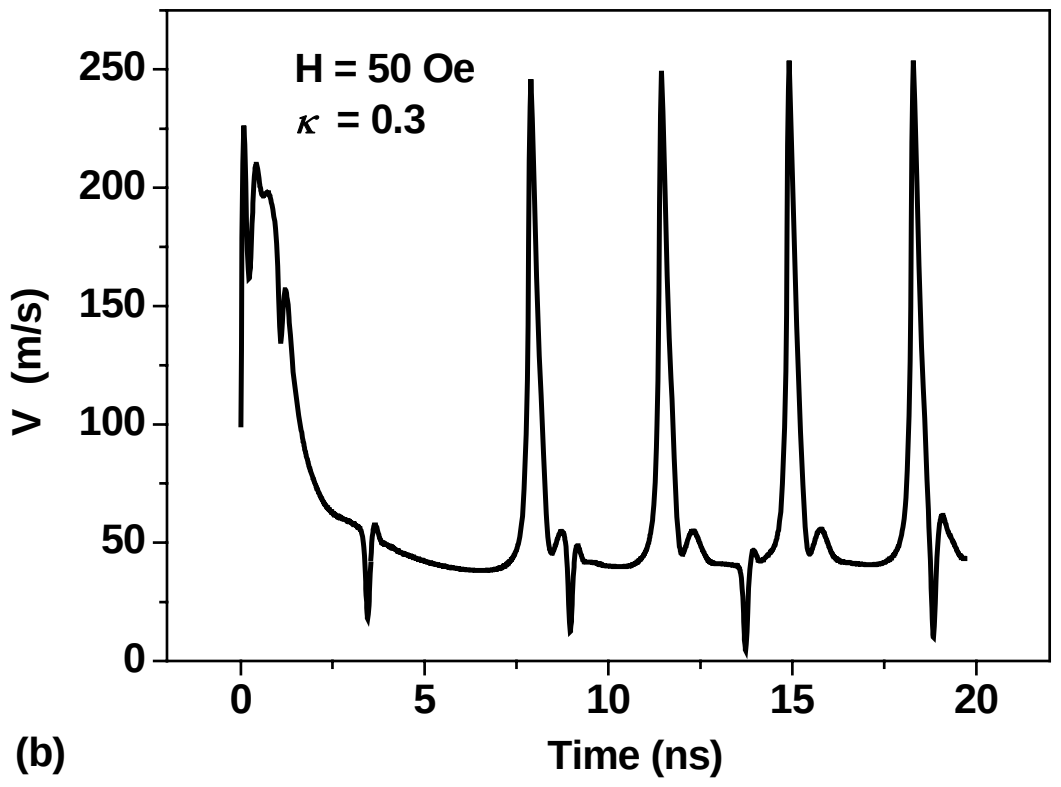

Fig. 4. a) The equilibrium magnetization distribution at the surface of the nanotube with $K > 0$ in the presence of a head-to-head domain wall located at the left edge of the tube; b) the domain wall velocity as a function of time in a applied magnetic field $H_z = 50$ Oe.

The authors wish to acknowledge the financial support of the Ministry of Education and Science of the Russian Federation in the framework of Increase Competitiveness Program of NUST «MISIS», contract № K2-2015-018.

## References

[1]W. F. Brown, Jr., *Micromagnetics* (Wiley, New York, 1963).
[2]M. T. Johnson, P. J. H. Bloemen, F. J. A. den Broeder and J. J. de Vries, Rep. Prog. Phys. **59**, 1409 (1996).
[3]H.N. Bertram and D.I. Paul, J. Appl. Phys. **82**, 2439 (1997).
[4]C. A. F. Vaz, J. A. C. Bland and G. Lauhoff, Rep. Prog. Phys. **71**, 056501 (2008).
[5]P.J. Jensen and K.H. Bennemann, Surf. Sci. Rep. **61**, 129 (2006).
[6]N.A. Usov and Yu. B. Grebenshchikov, J. Appl. Phys. **104,** 043903 (2008).
[7]P. Zhou, L. Zhang, and L. Deng, Appl. Phys. Lett. **96**, 112510 (2010).
[8]C. T. Sousa, D. C. Leitao, M. P. Proenca, J. Ventura, A. M. Pereira, and J. P. Araujo, Applied Physics Reviews **1,** 031102 (2014).
[9]C. Bran, E. M. Palmero, Zi-An Li, R. P. del Real, M. Spasova, M. Farle and M. Vazquez, J. Phys. D: Appl. Phys. **48,** 145304 (2015).
[10]J. Garcia, V.M. Prida, V. Vega, W.O. Rosa, R. Caballero-Flores, L. Iglesias, B. Hernando, J. Magn. Magn. Mater. **383**, 88 (2015).
[11]S. S. P. Parkin, M. Hayashi, and L. Thomas, Science **320**, 190 (2008).
[12]N. A. Usov, A. Zhukov and J. González, J. Magn. Magn. Mater. **316**, 255 (2007).
[13]J. S. Liu, F. X. Qin, D. M. Chen, H. X. Shen, H. Wang, D. W. Xing, M.-H. Phan, and J. F. Sun, J. Appl. Phys. **115**, 17A326 (2014).
[14]D. M. Chen, D. W. Xing, F. X. Qin, J. S. Liu, H. X. Shen, H. X. Peng, H. Wang, and J. F. Sun, J. Appl. Phys. **116**, 053907 (2014).